\documentclass[aps,prl,superscriptaddress,twocolumn]{revtex4-2}
\usepackage[final]{pdfpages}
\usepackage{graphicx}
\usepackage[utf8]{inputenc}
\usepackage{color}
\usepackage{amsmath,amsfonts,amssymb}
\usepackage{lpic}
\usepackage{ctable}
\usepackage{tabularx}
\usepackage{hyperref}

\newcommand{\Eqref}[1]{Eq.~(\ref{#1})}
\newcommand{\Figref}[1]{Fig.~\ref{#1}}

\newcommand{\dipc}{Donostia International Physics Center (DIPC), E-20018, Donostia-San Sebasti\'an, Spain}
\newcommand{\iker}{IKERBASQUE, Basque Foundation for Science, E-48013, Bilbao, Spain}
\newcommand{\cfm}{Centro de F\'{\i}sica de Materiales (CFM) CSIC-UPV/EHU, E-20018, Donostia-San~Sebasti\'an, Spain}
\newcommand{\dtu}{Department of Physics, Technical University of Denmark, DK-2800 Kgs.~Lyngby, Denmark}
\newcommand{\dtucomp}{DTU Computing Center, Technical University of Denmark, DK-2800 Kgs.~Lyngby, Denmark}
\newcommand{\cng}{Center for Nanostructured Graphene (CNG), Denmark}

\newcommand{\revision}[1]{\textcolor{black}{#1}}
\newcommand{\revisionB}[1]{\textcolor{black}{#1}}

\newcommand{\ie}{\emph{i.e.}}

\newcommand{\SIbands}{Figs.~S3 and S4 \cite{SM}}
\newcommand{\SIspinconfs}{Figs.~S6 and S7 \cite{SM}}
\newcommand{\SIindependentScatterers}{Sec.~S9 \cite{SM}}
\newcommand{\SIrotation}{Sec.~S11 \cite{SM}}
\newcommand{\SIbearded}{Sec.~S13 \cite{SM}}
\newcommand{\SIstatistics}{Sec.~S12 \cite{SM}}

\newcommand{\UPDN}{\fbox{$\uparrow\downarrow$}}
\newcommand{\UPUP}{\fbox{$\uparrow\uparrow$}}

\makeatletter
\patchcmd{\@outputpage@head}{\@ifx{\LS@rot\@undefined}{}{\LS@rot}}{}{}{}
\makeatother

\begin{document}

\title{Spin-polarizing electron beam splitter from crossed graphene nanoribbons}

\author{Sofia Sanz}
\email{sofia.sanz@dipc.org}
\affiliation{\dipc}

\author{Nick Papior}
\affiliation{\dtucomp}

\author{G\'eza Giedke}
\affiliation{\dipc}
\affiliation{\iker}

\author{Daniel S\'anchez-Portal}
\affiliation{\cfm}

\author{Mads Brandbyge}
\affiliation{\dtu}
\affiliation{\cng}

\author{Thomas~Frederiksen}
\email{thomas\_frederiksen@ehu.eus}
\affiliation{\dipc}
\affiliation{\iker}

\date{\today}

\begin{abstract}
Junctions composed of two crossed graphene nanoribbons (GNRs) have been theoretically proposed as electron beam splitters where incoming electron waves in one GNR can be split coherently into propagating waves in \emph{two} outgoing terminals with nearly equal amplitude and zero back-scattering.
Here we scrutinize this effect for devices composed of narrow zigzag GNRs taking explicitly into account the role of Coulomb repulsion that leads to spin-polarized edge states within mean-field theory.
We show that the beam-splitting effect survives the opening of the well-known correlation gap
and, more strikingly, that a \emph{spin-dependent} scattering potential emerges which spin-polarizes the transmitted electrons in the two outputs.
\revision{By studying different ribbons and intersection angles we provide evidence that this is a general feature with edge-polarized nanoribbons.}
A near-perfect polarization can be achieved by joining several junctions in series.
Our findings suggest that GNRs are interesting building blocks in spintronics and quantum technologies with applications for interferometry and entanglement.
\end{abstract}

\maketitle

\begin{figure*}
\centering
    \includegraphics[width=\textwidth]{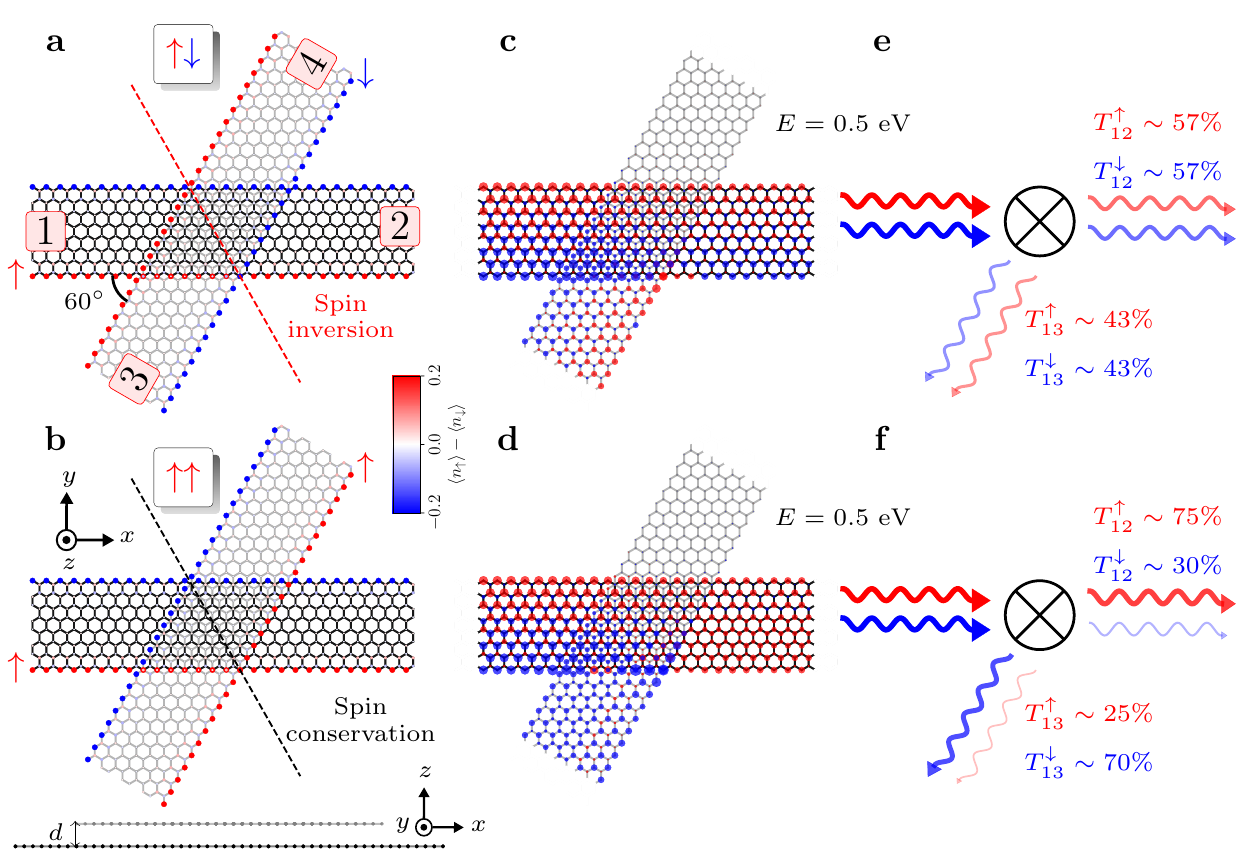}
    \caption{Transport setup and spin-dependent properties for AB-stacked 8-ZGNRs devices.
    (a,b) Two different self-consistent solutions for the spin-density distribution in the device region, labeled \UPDN\ and \UPUP, respectively, \revisionB{defined by the spin orientation of the lower edge of each GNR}.
    The up (down) spin density is shown in red (blue).
    The lower, horizontal ribbon is plotted in black, while the upper, intersecting at an angle of 60$^\circ$, is depicted in gray.
    Electrodes 1-4 are indicated. The ribbons are separated by a distance $d=3.34$ \AA \ along the $z$-axis, as displayed in the side view (lower part of b).
    The dashed lines in each configuration indicate a symmetry axis that maps the device geometry to itself through mirror operations, where the red (black) color of the axis further indicates that the spin index is inverted (conserved) by the symmetry operation.
    (c,d) Spin-resolved density of states of scattering states incoming from electrode 1 for the \UPDN\ and \UPUP\ spin configuration, respectively, computed at $E-E_F=0.5$ eV.
    The \revision{dominant} spin on each site at this energy is shown in red for up-spins and in blue for down-spins.
 (e,f) Sketch of incoming and outgoing waves through the scattering center (represented by the circled cross)  and the corresponding transmission probabilities from calculations.
}
    \label{fig:spin-confs}
\end{figure*}

Graphene is an exceptional material with attractive properties 
to explore fundamental physics and for use in technological applications \cite{CastroNeto2009}.
While ideal graphene is non-magnetic, custom-shaped
graphene nanostructures can be designed to exhibit complex magnetic phenomenology with promising possibilities for a new generation of nanoscale \textit{spintronics} devices \cite{Yazyev2010, Han2014}.
In fact, graphene $\pi$-magnetism is more delocalized and isotropic than conventional magnetism arising from $d$ or $f$ orbitals, which makes it electrically accessible \cite{Gonzalez-Herrero2016} and stable even at room temperature \cite{Magda2014}.
The intrinsically weak spin-orbit and hyperfine couplings in graphene lead to  
long spin coherence and relaxation times \cite{PhysRevLett.107.047207} as well as a long spin-diffusion length that is expected to reach $\sim 10\mu$m even at room temperature \cite{Tombros2007}.
This makes graphene an interesting platform for designing functionalities such as spin filters \cite{Son2006a,Hancock2010, SaFa.11.spinfilterdevice,GrPoJa.17.Nanostructuredgraphenespintronics}, spin qubits \cite{Trauzettel2007,PeFlPe.08.GrapheneAntidotLattices:} and electron quantum optics setups \cite{PhysRevLett.126.146803}.

Graphene nanoribbons (GNRs) have emerged as particularly attractive building blocks for molecular-scale electronic devices because they inherit some of the exceptional properties from graphene while having tunable electronic properties, such as the band gap dependency on their width and edge topology \cite{Son2006a}.
With the advent of bottom-up fabrication techniques, long defect-free samples of narrow GNRs can now be chemically produced via on-surface synthesis as demonstrated in the seminal works for armchair (AGNR) \cite{Cai2010} and zigzag (ZGNR) \cite{Ruffieux2016} ribbons.
Furthermore, manipulation of GNRs with scanning tunneling probes \cite{Koch2012, Kawai2016} opens the possibility to build two-dimensional multi-terminal graphene-based electronic circuits \cite{Jiao2010}, where their spin properties can be addressed by using spin polarized tips \cite{Wortmann2001} and probed by shot noise measurements \cite{Burtzlaff2015}.

Indeed, electron transport in GNR networks has been theoretically explored with the Landauer-B\"uttiker formalism \cite{Buettiker1985}
for a rich variety of multi-terminal device configurations \cite{Areshkin2007,Jayasekera2007,Botello-Mendez2011,CaCoMe.14.Electronicpropertiesthree}.
Most recently, crossed GNR junctions have been proposed as \emph{electron beam splitters} for 
electron quantum optics \cite{Lima2016, Brandimarte2017, Sanz2020}.
In these works it was found that by placing one GNR on top of another with a relative angle of $60^\circ$ the electron transfer process between the ribbons is strongly enhanced.
This enables to split incoming low-energy electron waves between two outgoing ports with a tunable ratio and negligible reflection probability,
an effect with roots in valley (chirality) preservation in the low-energy bands of ZGNRs \cite{Brey2006, Wakabayashi2007}.
However, since ZGNRs develop spin-polarized edge states, as theoretically \cite{Fujita1996} and experimentally \cite{Magda2014, BlZhBr.21.Spinsplittingdopant} demonstrated, one may expect that Coulomb repulsion could give rise to additional interesting features for the charge and spin transport in crossed ZGNRs.
\revision{For instance, it has been shown that the introduction of \emph{one} rough zigzag edge can be used to boost spin injection \cite{WiAdBe.08.SpinCurrentsRough}.}

In this manuscript we analyze the electronic structure and quantum transport properties of junctions composed of two infinite ZGNRs crossed with a relative angle of $60^{\circ}$
using the mean-field Hubbard (MFH) model in combination with nonequilibrium Green's functions (NEGF) to describe the open quantum systems \cite{dipc_hubbard}.
We show how the Coulomb repulsion opens a transport band gap and generates a spin-dependent scattering potential in the junction, which 
enables the devices to be operated as a \emph{spin-polarizing beam splitter}.

\begin{figure}[t]
    \centering
    \includegraphics[width=\columnwidth]{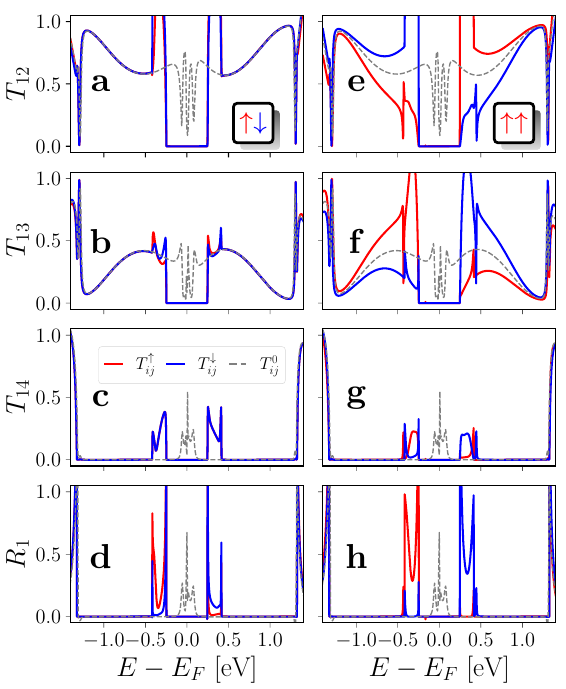}
    \caption{Spin- and energy-resolved transmission probabilities $T_{12}$, $T_{13}$, $T_{14}$ and 
    	reflection $R_1$ for (a-d) the \UPDN\ and (e-h) \UPUP\
    	configurations of \Figref{fig:spin-confs}.
    	Electrons are injected from electrode 1.
    	The red (blue) curves correspond to the up (down) spin components with $U=3$ eV.
    	For comparison, the corresponding calculations for the unpolarized case is 
    	indicated by dashed gray lines.
}
    \label{fig:spin-transmission}
\end{figure}

For a transparent analysis and efficient numerics we use the Hubbard Hamiltonian \cite{Hubbard1963} within the mean field approximation, well suited to describe $sp^2$ carbon systems \cite{Yazyev2010}, for both semi-infinite electrodes and device region as shown in \Figref{fig:spin-confs}, i.e.,
\begin{equation}\label{eq:Hubbard-Ham}
H_\mathrm{MFH} = \sum_{ij,\sigma}t_{ij}c^{\dagger}_{i\sigma}c_{j\sigma}^{\phantom{\dagger}} + U\sum_{i,\sigma}n_{i\sigma}\left\langle n_{i\overline{\sigma}}\right\rangle.
\end{equation}
Here $c_{i\sigma}$ is the annihilation operator of an electron at site $i$ with spin $\sigma=\{\uparrow,\downarrow\}$ and
$n_{i\sigma}=c_{i\sigma}^\dagger c_{i\sigma}^{\phantom{\dagger}}$ the corresponding number operator.
The matrix element $t_{ij}$ is computed by a two-center integral based on a Slater-Koster parametrization as explained in Ref.~\cite{Sanz2020} and
$U$ accounts for the Coulomb interaction between two electrons occupying the same $p_z$ orbital.
We fix $U=3$ eV which is in the typical range that yields a good agreement with \emph{ab initio} calculations
\cite{Yazyev2010,Hancock2010, GrPoJa.17.Nanostructuredgraphenespintronics, FePa.07.MagnetismGrapheneNanoislands,Li2019}.
The open system described by \Eqref{eq:Hubbard-Ham} is solved self-consistently using the NEGF method \cite{dipc_hubbard, Papior2017, zerothi_sisl} as detailed in the supplemental material (SM) \cite{SM}.
\revisionB{The corresponding many-electron state thus takes the form of a single Slater determinant of the occupied single-particle states from the MFH-NEGF equations.}

Figure \ref{fig:spin-confs}(a) shows the device structure for two AB-stacked ZGNRs, each with a width of 8 carbon atoms (8-ZGNRs).
In principle, away from the crossing (but within the spin correlation length), each of the four electrodes can be imposed one of the two possible symmetry-broken spin \revisionB{configurations at the edges}, leading to $2^4/2=8$ unique boundary conditions for the device region.
The self-consistent solutions to this problem are shown in \SIspinconfs\ for AB- and AA-stacked junctions, respectively, along with the electronic energy differences.
The spin polarization for the two lowest-energy \revisionB{states} with AB-stacking are shown in \Figref{fig:spin-confs}(a-b).
\revisionB{In the following we label these as \UPDN\ and \UPUP, where the first (second) arrow refers to the spin orientation of the lower edge of the horizontal (inclined) GNR.
Although the electronic energy of \UPUP\ is found to be 82 meV above that of \UPDN\ with AB-stacking,
it is interesting to consider both configurations as this (constant) energy penalty may be compensated by a (length-dependent) energy preference for a certain polarization on the extended GNRs through interactions with their environment.}

The spin- and energy-resolved transmission probability between any pair of electrodes can be computed from
$T^{\sigma}_{\alpha\beta}
	= \mathrm{Tr}\left[\mathbf{G}_{\sigma}\mathbf{\Gamma}_{\alpha \sigma} \mathbf{G}^{\dagger}_{\sigma}\mathbf{\Gamma}_{\beta \sigma} \right]$,
where $\mathbf{G}_{\sigma}$ is the device Green's function
and $\mathbf{\Gamma}_{\alpha  \sigma}
	= i(\mathbf{\Sigma}_{\alpha \sigma} - \mathbf{\Sigma}^{\dagger}_{\alpha \sigma})$
the broadening matrix related to the self-energy $\mathbf{\Sigma}_{\alpha \sigma}$ from electrode $\alpha$ and for spin-orientation $\sigma$ \cite{Buettiker1985, Papior2017}.
Similarly, the site-resolved density of scattering states can be computed as
$\mathbf{A}_{\alpha \sigma} = \mathbf{G}_{\sigma}\mathbf{\Gamma}_{\alpha  \sigma}\mathbf{G}^{\dagger}_{\sigma}$.

Figures \ref{fig:spin-confs}(c-d) show the spatial distribution of the scattering states incoming from electrode 1 in the conduction band.
At each lattice site the disk size is proportional to the density of states (summed over spin) while its color indicates the local majority spin.
The electron energy is chosen at $E= 0.5$ eV above the Fermi energy $E_F=0$, \revision{\ie, slightly away from the window with edge states.}
This implies mode propagation involving only a single GNR subband, cf.~\SIbands,
\revision{as well as robustness against edge disorder \cite{Zarbo2007}.}
%
Figures \ref{fig:spin-confs}(c-d) also illustrate how the transmitted wave---for both spin configurations \UPDN\ and \UPUP\ ---is split into electrodes 2 and 3 with negligible reflection and amplitude in electrode 4, as expected for the beam splitter.
Conceptually, this is expressed with the representation in \Figref{fig:spin-confs}(e-f)
along with the computed transmission probabilities. 

Remarkably, \UPDN\ and \UPUP\ differ substantially when
one considers the spin-resolved transmissions.
Whereas \UPDN\ does not polarize the current, since the transmission probabilities for both spin channels are equal, the \UPUP\ configuration leads to a ratio $T^{\downarrow}_{12}/T^{\uparrow}_{12} = 0.4$, i.e., a spin filtering effect.

\begin{figure}[t]
	\includegraphics[width=.8\columnwidth]{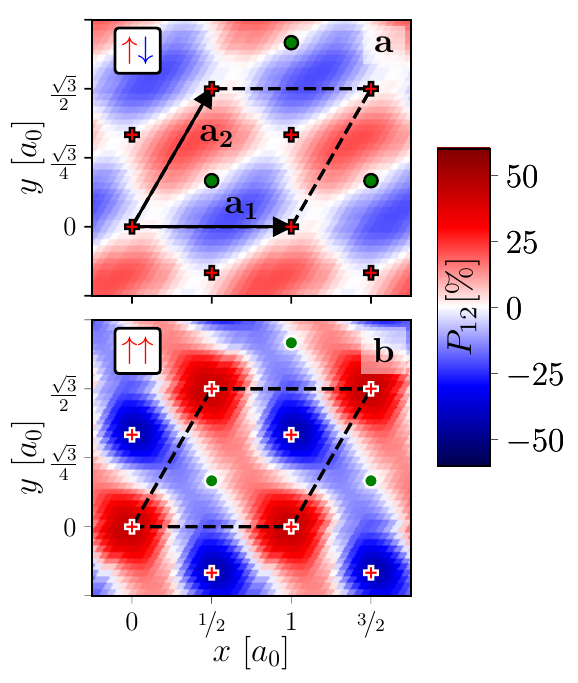}
\caption{Spin polarization $P_{12}$ of the current from electrode 1 to 2 as function of
	in-plane translations of one ribbon with respect to the other for (a) \UPUP\
	and (b) \UPDN\ configurations introduced in \Figref{fig:spin-confs}.
	The electron energy is in the conduction band at $E = 0.5$ eV.	 
	The in-plane unit cell (dashed lines) has lattice vectors $\mathbf{a}_1$ and $\mathbf{a}_2$, where \revision{$a_0 = 2.46$ {\AA}} is the graphene lattice constant.
	The red crosses (green disks) indicate the high-symmetry configurations with AB (AA) stacking.
}
	\label{fig:pol-translation}
\end{figure}

%
For further quantitative analysis, \Figref{fig:spin-transmission} reports the spin- and energy-resolved transmission and reflection probabilities for an electron injected from terminal $1$ into the \UPDN\ (panels a-d) and \UPUP\ (panels e-h) configurations.
For comparison, each panel includes the corresponding results for the unpolarized device ($U=0$, dashed gray lines) reported previously \cite{Sanz2020}.
The introduction of Coulomb repulsion has two direct consequences:
(i) it opens a transport gap near zero energy due to polarization of the edge bands, and
(ii) it shifts the states at the Brillouin zone boundary (\SIbands) resulting in the formation of \emph{two} transverse modes at very low energy.
While the beam splitting effect in the two-mode energy range is hampered by substantial scattering and reflection (\Figref{fig:spin-transmission}d,h),
it is completely restored in the energy range with only a single mode, i.e., 0.4 eV $< |E| <$ 1.3 eV, a condition already identified for unpolarized devices \cite{Sanz2020}.
In fact, the transmission properties for \UPDN\ coincides there with those of the unpolarized device (\Figref{fig:spin-transmission}a-d).
On the other hand, for the \UPUP\ configuration the probabilities $T_{12}$ and $T_{13}$ show a strong spin splitting (\Figref{fig:spin-transmission}e-h),
revealing that the spin-filtering effect emphasized in \Figref{fig:spin-confs}(d,f) exists for the whole band.

This qualitative difference between \UPDN\ and \UPUP\ can be understood by considering
the different symmetries that apply to these two configurations.
Geometrically, the considered AB-stacked structure possesses one mirror-symmetry plane as shown by the dashed lines in \Figref{fig:spin-confs}(a,b) \cite{Sanz2020}.
The difference emerges when one considers symmetry-lowering by the spin polarization:
For \UPDN\ the spin index maps into the opposite through the mirror operation (red axis) while for \UPUP\ the spin index is conserved.
More specifically for \UPDN, these spatial symmetries impose constraints in the transmission probabilities \emph{between} the spin channels, e.g.,~that $T^{\sigma}_{12}=T^{\Bar{\sigma}}_{43}$, $T^{\sigma}_{13}=T^{\Bar{\sigma}}_{42}$, etc.
Further, considering probability conservation for injection from electrodes 1 or 2 one has the relations $T^{\sigma}_{12}+T^{\sigma}_{13}=T^{\bar{\sigma}}_{21}+T^{\bar{\sigma}}_{24}=1$ (valid when $R^{\sigma}_{1}=T^{\sigma}_{14}=R^{\Bar{\sigma}}_{2}=T^{\Bar{\sigma}}_{23}=0$). 
Together with time-reversal symmetry ($T^\sigma_{ij}=T^\sigma_{ji}$) it follows that $T^{\sigma}_{12}=T^{\Bar{\sigma}}_{12}$ in the case of \UPDN,
i.e., that the transmissions are spin \emph{independent}.
For \UPUP\ no such condition applies and the spin channels are decoupled and the transmission probabilities may be very different.
Indeed, this is directly seen in our calculations.

\begin{figure}[t]
	\includegraphics[width=\columnwidth]{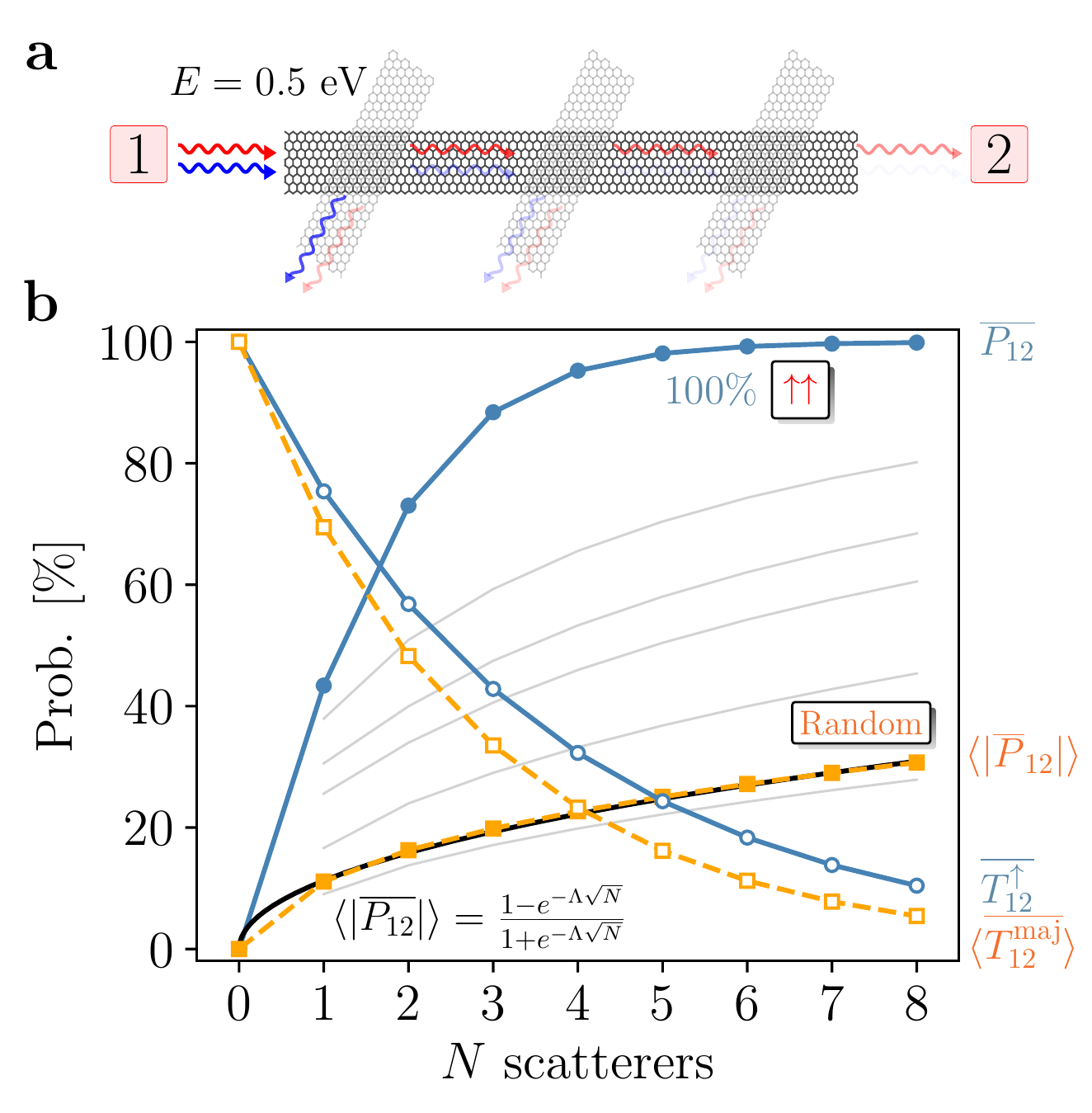}
\caption{(a) Sketch of an array of three consecutive AB-stacked 8-ZGNRs crossings to enhance the spin-polarized current at the output electrode 2.
(b) Spin polarization \revision{$\overline{P_{12}}(N)$ (filled symbols) and majority-spin transmission $\overline{T^\mathrm{maj}_{12}}(N)$ (open symbols)} as a function of the number of crossings $N$ between terminals 1 and 2 in the conduction band at $E = 0.5$ eV.
\revision{Two different scenarios are considered:
an ideal arrangement of identical \UPUP\ AB-crossings (blue circles) as well as a random sampling (orange squares) over $10^7$ different spin-, intersection angle (within 55-65$^\circ$), and translation configurations drawn from the data in \SIrotation\ assuming equal weights.
The average polarization $\langle |\overline{P_{12}}(N)|\rangle$ follows an analytic expression (black line, see Sec.~S12 \cite{SM}) approaching 1 exponentially in $\sqrt{N}$.
The gray lines indicate the best (1, 5, 10, 25, 50)th percentiles of the random distribution.
}
}
\label{fig:crossings-array}
\end{figure}

If we consider junction imperfections the aforementioned symmetry constraint
would be absent and the spin-polarizing effect no longer symmetry forbidden.
To examine the relationship between geometry and transport properties we use
as a measure the spin polarization in the transmission between a pair of electrodes:
\begin{equation}\label{eq:polarization}
P_{\alpha\beta}
	= \frac{T^{\uparrow}_{\alpha\beta}-T^{\downarrow}_{\alpha\beta}}
	{T^{\uparrow}_{\alpha\beta}+T^{\downarrow}_{\alpha\beta}}.
\end{equation}
Figure \ref{fig:pol-translation} shows $P_{12}$ at $E = 0.5$ eV
as a function of in-plane translations of one ribbon with respect to the other for 
both \UPUP\ and \UPDN\ configurations.
The AB- and AA-stacked geometries are indicated with symbols in the density plots.
Evidently, away from these high-symmetry situations the spin-polarizing effect is generally present.
\revision{The same conclusion holds true also for a range of twist angles (\SIrotation).}

At this point it should be noted that it may be difficult to prepare the device in one specific spin configuration, such as the low-energy states \UPDN\ and \UPUP\ discussed up to now.
For instance, it is not possible to tune which one is the energetically lower (and thus at low temperatures thermally stable) state by a homogeneous magnetic field as the Zeeman energy is the same for both solutions.
On the other hand, transverse electric fields across the individual electrodes \cite{Son2006a}
or injection of spin-polarized currents at the edges from the tip of an STM \cite{LoBeTe.10.Controllingstateof} could potentially be strategies
to control their magnetization.
Nevertheless, our fundamental assumption is that the different collective
spin states of the device are sufficiently long lived and robust to be probed by
a transient current pulse.
This assumption is supported by the fact that our calculations predict that the electronic
energy is increased by about 0.20 eV when a magnetic domain wall is inserted into a 8-ZGNR
(\SIspinconfs), an indication of a very large barrier even compared to room temperature.

The spin-polarizing effect of a single junction discussed above can be enhanced 
by placing several consecutive crossings to form an array of scatterers as displayed in \Figref{fig:crossings-array}(a).
Because back-scattering is negligible in the single-mode energy region, we can approximate the overall transmission probability across an array of $N$ crossings as
$\overline{T^{\sigma}_{12}} \approx \prod_i^N {T^{\sigma(i)}_{12}}$
where ${T^{\sigma(i)}_{12}}$ is the transmission of the $i$th junction.
This approximation was tested for the case of $N=3$ and shows an excellent agreement compared to a calculation of the full device (\revision{\SIindependentScatterers}).
This idea is exemplified in \Figref{fig:crossings-array}(b) \revision{for two different scenarios: \revisionB{an ideal arrangement of} identical \UPUP\ AB-stacked configurations (blue circles) 
as well as a \revisionB{more realistic situation corresponding to} random sampling (orange squares, \SIstatistics) over different spin-, intersection angle (within 55-65$^\circ$), and translation configurations.}
This shows that with four crossings the total current polarization can reach $\overline{P_{12}}\sim 95\%$ with a transmission of $\overline{T^\uparrow_{12}} \sim 32 \%$ \revision{in the ideal case.
	Even in the pessimistic case with random junctions, where partial cancellation can occur due to sign changes in the individual $P_{12}$, the
spin polarization $\langle|\overline{P_{12}}|\rangle$ of the array approaches 1 exponentially in $\sqrt{N}$ (black curve, see \cite{SM}).
The best 1st percentile (top gray curve) of the sampled arrays still reaches $\overline{P_{12}}\sim 80\%$ for $N=8$.
}
\revisionB{Although this statistical analysis---described in detail in \SIstatistics---is based on the simplifying assumption of equal weights of the configurations,
it serves to illustrate that arrays can be interesting even if one does not have precise control over the individual junctions.}

In conclusion, we have analyzed the spin-dependent transport properties of crossed ZGNRs using MFH and NEGF theory,
and found that the beam-splitting effect reported previously survives in presence of Coulomb repulsions with two distinct modifications: 
a transport gap opens at low energies and a \emph{spin-dependent} scattering potential emerges.
Except for specific high-symmetry configurations, this class of electronic devices are generally predicted to behave as \emph{spin-polarizing beam splitters} with interesting possibilities 
for electron quantum optics \cite{BoFrPa.14.Electronquantumoptics}.
\revision{Such spin-dependent scattering potentials are also obtained with other edge-polarized nanoribbons (\SIbearded).}
By constructing arrays of junctions the spin-polarizing effect can be enhanced.

\revision{Although the proposed devices are ahead of current experiments, a rapid progress in bottom-up fabrication and scanning probe techniques makes it conceivable to assemble nearly defect-free junctions on insulating thin films \cite{JaMaZe.18.MappingConductanceElectronically}, to drive coherent electron dynamics \cite{BaPaCh.15.Electronparamagneticresonance,AmJeWe.21.Lightwavedrivenscanning}, and to characterize electron transport by multi-probe setups \cite{KoBrLi.19.Electronictransportplanar} or through single-photon emission \cite{ChAfSc.18.BrightElectroluminescenceSingle}.
}
Our results add to the vision of using GNR-based devices for spintronics and quantum technologies.
For instance, two spin-polarizing beam splitters in combination with a charge detector can be used to deterministically entangle a \revision{moving} spin qubit \cite{BeDiEm.04.ChargeDetectionEnables}.
Conversely, a spin-polarizing beam splitter can also be used to determine the entanglement of injected pairs of spins \cite{MaBrRe.13.Spinfilteringentanglement}.
As an additional application, a high-fidelity spin filter allows "spin-to-charge" conversion and thus a charge-measurement-based spin determination.

\begin{acknowledgments}
\revision{This work was supported by the Spanish MCIN/AEI/
10.13039/501100011033 (PID2020-115406GB-I00 and PID2019-107338RB-C66), 
\revision{the Basque Department of Education (PRE-2021-2-0190 and PIBA-2020-1-0014)}, the University of the Basque Country (IT1246-19), and the European Union (EU) through Horizon 2020 (FET-Open project ``{SPRING}'' Grant no.~863098).}
\end{acknowledgments}

\nocite{FeMeLa.11.DynamicalSignaturesEdge, Brandbyge2002, Papior2017, Sancho1985, zerothi_sisl, Papior2017, Asano2019, Mostaani2015, Lee2005, Sanz2020, Cahay1988}

\bibliographystyle{apsrev-title}
\bibliography{main.bbl}

\begin{thebibliography}{58}
\makeatletter
\providecommand \@ifxundefined [1]{%
 \@ifx{#1\undefined}
}%
\providecommand \@ifnum [1]{%
 \ifnum #1\expandafter \@firstoftwo
 \else \expandafter \@secondoftwo
 \fi
}%
\providecommand \@ifx [1]{%
 \ifx #1\expandafter \@firstoftwo
 \else \expandafter \@secondoftwo
 \fi
}%
\providecommand \natexlab [1]{#1}%
\providecommand \enquote  [1]{``#1''}%
\providecommand \bibnamefont  [1]{#1}%
\providecommand \bibfnamefont [1]{#1}%
\providecommand \citenamefont [1]{#1}%
\providecommand \href@noop [0]{\@secondoftwo}%
\providecommand \href [0]{\begingroup \@sanitize@url \@href}%
\providecommand \@href[1]{\@@startlink{#1}\@@href}%
\providecommand \@@href[1]{\endgroup#1\@@endlink}%
\providecommand \@sanitize@url [0]{\catcode `\\12\catcode `\$12\catcode
  `\&12\catcode `\#12\catcode `\^12\catcode `\_12\catcode `\%12\relax}%
\providecommand \@@startlink[1]{}%
\providecommand \@@endlink[0]{}%
\providecommand \url  [0]{\begingroup\@sanitize@url \@url }%
\providecommand \@url [1]{\endgroup\@href {#1}{\urlprefix }}%
\providecommand \urlprefix  [0]{URL }%
\providecommand \Eprint [0]{\href }%
\providecommand \doibase [0]{https://doi.org/}%
\providecommand \selectlanguage [0]{\@gobble}%
\providecommand \bibinfo  [0]{\@secondoftwo}%
\providecommand \bibfield  [0]{\@secondoftwo}%
\providecommand \translation [1]{[#1]}%
\providecommand \BibitemOpen [0]{}%
\providecommand \bibitemStop [0]{}%
\providecommand \bibitemNoStop [0]{.\EOS\space}%
\providecommand \EOS [0]{\spacefactor3000\relax}%
\providecommand \BibitemShut  [1]{\csname bibitem#1\endcsname}%
\let\auto@bib@innerbib\@empty

\bibitem[{\citenamefont{Castro~Neto et~al.}(2009)\citenamefont{Castro~Neto,
  Guinea, Peres, Novoselov, and Geim}}]{CastroNeto2009}
\bibinfo{author}{\bibfnamefont{A.~H.} \bibnamefont{Castro~Neto}},
  \bibinfo{author}{\bibfnamefont{F.}~\bibnamefont{Guinea}},
  \bibinfo{author}{\bibfnamefont{N.~M.~R.} \bibnamefont{Peres}},
  \bibinfo{author}{\bibfnamefont{K.~S.} \bibnamefont{Novoselov}},
  \bibnamefont{and} \bibinfo{author}{\bibfnamefont{A.~K.} \bibnamefont{Geim}},
  ``The electronic properties of graphene,'' \bibinfo{journal}{Rev. Mod. Phys.}
  \textbf{\bibinfo{volume}{81}}, \bibinfo{pages}{109} (\bibinfo{year}{2009}).

\bibitem[{\citenamefont{Yazyev}(2010)}]{Yazyev2010}
\bibinfo{author}{\bibfnamefont{O.~V.} \bibnamefont{Yazyev}}, ``Emergence of
  magnetism in graphene materials and nanostructures,'' \bibinfo{journal}{Rep.
  Prog. Phys.} \textbf{\bibinfo{volume}{73}}, \bibinfo{pages}{056501}
  (\bibinfo{year}{2010}).

\bibitem[{\citenamefont{Han et~al.}(2014)\citenamefont{Han, Kawakami, Gmitra,
  and Fabian}}]{Han2014}
\bibinfo{author}{\bibfnamefont{W.}~\bibnamefont{Han}},
  \bibinfo{author}{\bibfnamefont{R.~K.} \bibnamefont{Kawakami}},
  \bibinfo{author}{\bibfnamefont{M.}~\bibnamefont{Gmitra}}, \bibnamefont{and}
  \bibinfo{author}{\bibfnamefont{J.}~\bibnamefont{Fabian}}, ``Graphene
  spintronics,'' \bibinfo{journal}{Nat.~Nanotechnol.}
  \textbf{\bibinfo{volume}{9}}, \bibinfo{pages}{794} (\bibinfo{year}{2014}).

\bibitem[{\citenamefont{Gonz{\'a}lez-Herrero
  et~al.}(2016)\citenamefont{Gonz{\'a}lez-Herrero, G{\'o}mez-Rodr{\'\i}guez,
  Mallet, Moaied, Palacios, Salgado, Ugeda, Veuillen, Yndurain, and
  Brihuega}}]{Gonzalez-Herrero2016}
\bibinfo{author}{\bibfnamefont{H.}~\bibnamefont{Gonz{\'a}lez-Herrero}},
  \bibinfo{author}{\bibfnamefont{J.~M.}
  \bibnamefont{G{\'o}mez-Rodr{\'\i}guez}},
  \bibinfo{author}{\bibfnamefont{P.}~\bibnamefont{Mallet}},
  \bibinfo{author}{\bibfnamefont{M.}~\bibnamefont{Moaied}},
  \bibinfo{author}{\bibfnamefont{J.~J.} \bibnamefont{Palacios}},
  \bibinfo{author}{\bibfnamefont{C.}~\bibnamefont{Salgado}},
  \bibinfo{author}{\bibfnamefont{M.~M.} \bibnamefont{Ugeda}},
  \bibinfo{author}{\bibfnamefont{J.-Y.} \bibnamefont{Veuillen}},
  \bibinfo{author}{\bibfnamefont{F.}~\bibnamefont{Yndurain}}, \bibnamefont{and}
  \bibinfo{author}{\bibfnamefont{I.}~\bibnamefont{Brihuega}}, ``Atomic-scale
  control of graphene magnetism by using hydrogen atoms,''
  \bibinfo{journal}{Science} \textbf{\bibinfo{volume}{352}},
  \bibinfo{pages}{437} (\bibinfo{year}{2016}).

\bibitem[{\citenamefont{Magda et~al.}(2014)\citenamefont{Magda, Jin,
  Hagym{\'a}si, Vancs{\'o}, Osv{\'a}th, Nemes-Incze, Hwang, Bir{\'o}, and
  Tapaszt{\'o}}}]{Magda2014}
\bibinfo{author}{\bibfnamefont{G.~Z.} \bibnamefont{Magda}},
  \bibinfo{author}{\bibfnamefont{X.}~\bibnamefont{Jin}},
  \bibinfo{author}{\bibfnamefont{I.}~\bibnamefont{Hagym{\'a}si}},
  \bibinfo{author}{\bibfnamefont{P.}~\bibnamefont{Vancs{\'o}}},
  \bibinfo{author}{\bibfnamefont{Z.}~\bibnamefont{Osv{\'a}th}},
  \bibinfo{author}{\bibfnamefont{P.}~\bibnamefont{Nemes-Incze}},
  \bibinfo{author}{\bibfnamefont{C.}~\bibnamefont{Hwang}},
  \bibinfo{author}{\bibfnamefont{L.~P.} \bibnamefont{Bir{\'o}}},
  \bibnamefont{and}
  \bibinfo{author}{\bibfnamefont{L.}~\bibnamefont{Tapaszt{\'o}}},
  ``Room-temperature magnetic order on zigzag edges of narrow graphene
  nanoribbons,'' \bibinfo{journal}{Nature} \textbf{\bibinfo{volume}{514}},
  \bibinfo{pages}{608 EP } (\bibinfo{year}{2014}).

\bibitem[{\citenamefont{Han and Kawakami}(2011)}]{PhysRevLett.107.047207}
\bibinfo{author}{\bibfnamefont{W.}~\bibnamefont{Han}} \bibnamefont{and}
  \bibinfo{author}{\bibfnamefont{R.~K.} \bibnamefont{Kawakami}}, ``Spin
  relaxation in single-layer and bilayer graphene,'' \bibinfo{journal}{Phys.
  Rev. Lett.} \textbf{\bibinfo{volume}{107}}, \bibinfo{pages}{047207}
  (\bibinfo{year}{2011}).

\bibitem[{\citenamefont{Tombros et~al.}(2007)\citenamefont{Tombros, Jozsa,
  Popinciuc, Jonkman, and van Wees}}]{Tombros2007}
\bibinfo{author}{\bibfnamefont{N.}~\bibnamefont{Tombros}},
  \bibinfo{author}{\bibfnamefont{C.}~\bibnamefont{Jozsa}},
  \bibinfo{author}{\bibfnamefont{M.}~\bibnamefont{Popinciuc}},
  \bibinfo{author}{\bibfnamefont{H.~T.} \bibnamefont{Jonkman}},
  \bibnamefont{and} \bibinfo{author}{\bibfnamefont{B.~J.} \bibnamefont{van
  Wees}}, ``Electronic spin transport and spin precession in single graphene
  layers at room temperature,'' \bibinfo{journal}{Nature}
  \textbf{\bibinfo{volume}{448}}, \bibinfo{pages}{571} (\bibinfo{year}{2007}).

\bibitem[{\citenamefont{Son et~al.}(2006)\citenamefont{Son, Cohen, and
  Louie}}]{Son2006a}
\bibinfo{author}{\bibfnamefont{Y.-W.} \bibnamefont{Son}},
  \bibinfo{author}{\bibfnamefont{M.~L.} \bibnamefont{Cohen}}, \bibnamefont{and}
  \bibinfo{author}{\bibfnamefont{S.~G.} \bibnamefont{Louie}}, ``Half-metallic
  graphene nanoribbons,'' \bibinfo{journal}{Nature}
  \textbf{\bibinfo{volume}{444}}, \bibinfo{pages}{347} (\bibinfo{year}{2006}).

\bibitem[{\citenamefont{Hancock et~al.}(2010)\citenamefont{Hancock, Uppstu,
  Saloriutta, Harju, and Puska}}]{Hancock2010}
\bibinfo{author}{\bibfnamefont{Y.}~\bibnamefont{Hancock}},
  \bibinfo{author}{\bibfnamefont{A.}~\bibnamefont{Uppstu}},
  \bibinfo{author}{\bibfnamefont{K.}~\bibnamefont{Saloriutta}},
  \bibinfo{author}{\bibfnamefont{A.}~\bibnamefont{Harju}}, \bibnamefont{and}
  \bibinfo{author}{\bibfnamefont{M.~J.} \bibnamefont{Puska}}, ``Generalized
  tight-binding transport model for graphene nanoribbon-based systems,''
  \bibinfo{journal}{Phys. Rev. B} \textbf{\bibinfo{volume}{81}},
  \bibinfo{pages}{245402} (\bibinfo{year}{2010}).

\bibitem[{\citenamefont{Saffarzadeh and
  Farghadan}(2011)}]{SaFa.11.spinfilterdevice}
\bibinfo{author}{\bibfnamefont{A.}~\bibnamefont{Saffarzadeh}} \bibnamefont{and}
  \bibinfo{author}{\bibfnamefont{R.}~\bibnamefont{Farghadan}}, ``A spin-filter
  device based on armchair graphene nanoribbons,'' \bibinfo{journal}{Appl.
  Phys. Lett.} \textbf{\bibinfo{volume}{98}}, \bibinfo{pages}{023106}
  (\bibinfo{year}{2011}).

\bibitem[{\citenamefont{Gregersen et~al.}(2017)\citenamefont{Gregersen, Power,
  and Jauho}}]{GrPoJa.17.Nanostructuredgraphenespintronics}
\bibinfo{author}{\bibfnamefont{S.~S.} \bibnamefont{Gregersen}},
  \bibinfo{author}{\bibfnamefont{S.~R.} \bibnamefont{Power}}, \bibnamefont{and}
  \bibinfo{author}{\bibfnamefont{A.-P.} \bibnamefont{Jauho}}, ``Nanostructured
  graphene for spintronics,'' \bibinfo{journal}{Phys. Rev. B}
  \textbf{\bibinfo{volume}{95}}, \bibinfo{pages}{121406}
  (\bibinfo{year}{2017}).

\bibitem[{\citenamefont{Trauzettel et~al.}(2007)\citenamefont{Trauzettel,
  Bulaev, Loss, and Burkard}}]{Trauzettel2007}
\bibinfo{author}{\bibfnamefont{B.}~\bibnamefont{Trauzettel}},
  \bibinfo{author}{\bibfnamefont{D.~V.} \bibnamefont{Bulaev}},
  \bibinfo{author}{\bibfnamefont{D.}~\bibnamefont{Loss}}, \bibnamefont{and}
  \bibinfo{author}{\bibfnamefont{G.}~\bibnamefont{Burkard}}, ``{Spin qubits in
  graphene quantum dots},'' \bibinfo{journal}{Nat. Phys.}
  \textbf{\bibinfo{volume}{3}}, \bibinfo{pages}{192} (\bibinfo{year}{2007}).

\bibitem[{\citenamefont{Pedersen et~al.}(2008)\citenamefont{Pedersen, Flindt,
  Pedersen, Mortensen, Jauho, and
  Pedersen}}]{PeFlPe.08.GrapheneAntidotLattices:}
\bibinfo{author}{\bibfnamefont{T.~G.} \bibnamefont{Pedersen}},
  \bibinfo{author}{\bibfnamefont{C.}~\bibnamefont{Flindt}},
  \bibinfo{author}{\bibfnamefont{J.}~\bibnamefont{Pedersen}},
  \bibinfo{author}{\bibfnamefont{N.~A.} \bibnamefont{Mortensen}},
  \bibinfo{author}{\bibfnamefont{A.-P.} \bibnamefont{Jauho}}, \bibnamefont{and}
  \bibinfo{author}{\bibfnamefont{K.}~\bibnamefont{Pedersen}}, ``Graphene
  antidot lattices: Designed defects and spin qubits,''
  \bibinfo{journal}{Phys.~Rev.~Lett.} \textbf{\bibinfo{volume}{100}},
  \bibinfo{pages}{136804} (\bibinfo{year}{2008}).

\bibitem[{\citenamefont{Jo et~al.}(2021)\citenamefont{Jo, Brasseur, Assouline,
  Fleury, Sim, Watanabe, Taniguchi, Dumnernpanich, Roche, Glattli
  et~al.}}]{PhysRevLett.126.146803}
\bibinfo{author}{\bibfnamefont{M.}~\bibnamefont{Jo}},
  \bibinfo{author}{\bibfnamefont{P.}~\bibnamefont{Brasseur}},
  \bibinfo{author}{\bibfnamefont{A.}~\bibnamefont{Assouline}},
  \bibinfo{author}{\bibfnamefont{G.}~\bibnamefont{Fleury}},
  \bibinfo{author}{\bibfnamefont{H.-S.} \bibnamefont{Sim}},
  \bibinfo{author}{\bibfnamefont{K.}~\bibnamefont{Watanabe}},
  \bibinfo{author}{\bibfnamefont{T.}~\bibnamefont{Taniguchi}},
  \bibinfo{author}{\bibfnamefont{W.}~\bibnamefont{Dumnernpanich}},
  \bibinfo{author}{\bibfnamefont{P.}~\bibnamefont{Roche}},
  \bibinfo{author}{\bibfnamefont{D.~C.} \bibnamefont{Glattli}},
  \bibnamefont{et~al.}, ``Quantum {Hall} valley splitters and a tunable
  {Mach-Zehnder} interferometer in graphene,'' \bibinfo{journal}{Phys. Rev.
  Lett.} \textbf{\bibinfo{volume}{126}}, \bibinfo{pages}{146803}
  (\bibinfo{year}{2021}).

\bibitem[{\citenamefont{Cai et~al.}(2010)\citenamefont{Cai, Ruffieux, Jaafar,
  Bieri, Braun, Blankenburg, Muoth, Seitsonen, Saleh, Feng et~al.}}]{Cai2010}
\bibinfo{author}{\bibfnamefont{J.}~\bibnamefont{Cai}},
  \bibinfo{author}{\bibfnamefont{P.}~\bibnamefont{Ruffieux}},
  \bibinfo{author}{\bibfnamefont{R.}~\bibnamefont{Jaafar}},
  \bibinfo{author}{\bibfnamefont{M.}~\bibnamefont{Bieri}},
  \bibinfo{author}{\bibfnamefont{T.}~\bibnamefont{Braun}},
  \bibinfo{author}{\bibfnamefont{S.}~\bibnamefont{Blankenburg}},
  \bibinfo{author}{\bibfnamefont{M.}~\bibnamefont{Muoth}},
  \bibinfo{author}{\bibfnamefont{A.~P.} \bibnamefont{Seitsonen}},
  \bibinfo{author}{\bibfnamefont{M.}~\bibnamefont{Saleh}},
  \bibinfo{author}{\bibfnamefont{X.}~\bibnamefont{Feng}}, \bibnamefont{et~al.},
  ``Atomically precise bottom-up fabrication of graphene nanoribbons,''
  \bibinfo{journal}{Nature} \textbf{\bibinfo{volume}{466}},
  \bibinfo{pages}{470} (\bibinfo{year}{2010}).

\bibitem[{\citenamefont{Ruffieux et~al.}(2016)\citenamefont{Ruffieux, Wang,
  Yang, Sánchez-Sánchez, Liu, Dienel, Talirz, Shinde, Pignedoli, Passerone
  et~al.}}]{Ruffieux2016}
\bibinfo{author}{\bibfnamefont{P.}~\bibnamefont{Ruffieux}},
  \bibinfo{author}{\bibfnamefont{S.}~\bibnamefont{Wang}},
  \bibinfo{author}{\bibfnamefont{B.}~\bibnamefont{Yang}},
  \bibinfo{author}{\bibfnamefont{C.}~\bibnamefont{Sánchez-Sánchez}},
  \bibinfo{author}{\bibfnamefont{J.}~\bibnamefont{Liu}},
  \bibinfo{author}{\bibfnamefont{T.}~\bibnamefont{Dienel}},
  \bibinfo{author}{\bibfnamefont{L.}~\bibnamefont{Talirz}},
  \bibinfo{author}{\bibfnamefont{P.}~\bibnamefont{Shinde}},
  \bibinfo{author}{\bibfnamefont{C.~A.} \bibnamefont{Pignedoli}},
  \bibinfo{author}{\bibfnamefont{D.}~\bibnamefont{Passerone}},
  \bibnamefont{et~al.}, ``On-surface synthesis of graphene nanoribbons with
  zigzag edge topology,'' \bibinfo{journal}{Nature}
  \textbf{\bibinfo{volume}{531}}, \bibinfo{pages}{489} (\bibinfo{year}{2016}).

\bibitem[{\citenamefont{Koch et~al.}(2012)\citenamefont{Koch, Ample, Joachim,
  and Grill}}]{Koch2012}
\bibinfo{author}{\bibfnamefont{M.}~\bibnamefont{Koch}},
  \bibinfo{author}{\bibfnamefont{F.}~\bibnamefont{Ample}},
  \bibinfo{author}{\bibfnamefont{C.}~\bibnamefont{Joachim}}, \bibnamefont{and}
  \bibinfo{author}{\bibfnamefont{L.}~\bibnamefont{Grill}}, ``Voltage-dependent
  conductance of a single graphene nanoribbon,''
  \bibinfo{journal}{Nat.~Nanotechnol.} \textbf{\bibinfo{volume}{7}},
  \bibinfo{pages}{713} (\bibinfo{year}{2012}).

\bibitem[{\citenamefont{Kawai et~al.}(2016)\citenamefont{Kawai, Benassi,
  Gnecco, S{\"o}de, Pawlak, Feng, M{\"u}llen, Passerone, Pignedoli, Ruffieux
  et~al.}}]{Kawai2016}
\bibinfo{author}{\bibfnamefont{S.}~\bibnamefont{Kawai}},
  \bibinfo{author}{\bibfnamefont{A.}~\bibnamefont{Benassi}},
  \bibinfo{author}{\bibfnamefont{E.}~\bibnamefont{Gnecco}},
  \bibinfo{author}{\bibfnamefont{H.}~\bibnamefont{S{\"o}de}},
  \bibinfo{author}{\bibfnamefont{R.}~\bibnamefont{Pawlak}},
  \bibinfo{author}{\bibfnamefont{X.}~\bibnamefont{Feng}},
  \bibinfo{author}{\bibfnamefont{K.}~\bibnamefont{M{\"u}llen}},
  \bibinfo{author}{\bibfnamefont{D.}~\bibnamefont{Passerone}},
  \bibinfo{author}{\bibfnamefont{C.~A.} \bibnamefont{Pignedoli}},
  \bibinfo{author}{\bibfnamefont{P.}~\bibnamefont{Ruffieux}},
  \bibnamefont{et~al.}, ``Superlubricity of graphene nanoribbons on gold
  surfaces,'' \bibinfo{journal}{Science} \textbf{\bibinfo{volume}{351}},
  \bibinfo{pages}{957} (\bibinfo{year}{2016}).

\bibitem[{\citenamefont{Jiao et~al.}(2010)\citenamefont{Jiao, Zhang, Ding, Liu,
  and Dai}}]{Jiao2010}
\bibinfo{author}{\bibfnamefont{L.}~\bibnamefont{Jiao}},
  \bibinfo{author}{\bibfnamefont{L.}~\bibnamefont{Zhang}},
  \bibinfo{author}{\bibfnamefont{L.}~\bibnamefont{Ding}},
  \bibinfo{author}{\bibfnamefont{J.}~\bibnamefont{Liu}}, \bibnamefont{and}
  \bibinfo{author}{\bibfnamefont{H.}~\bibnamefont{Dai}}, ``Aligned graphene
  nanoribbons and crossbars from unzipped carbon nanotubes,''
  \bibinfo{journal}{Nano Research} \textbf{\bibinfo{volume}{3}},
  \bibinfo{pages}{387} (\bibinfo{year}{2010}).

\bibitem[{\citenamefont{Wortmann et~al.}(2001)\citenamefont{Wortmann, Heinze,
  Kurz, Bihlmayer, and Bl\"ugel}}]{Wortmann2001}
\bibinfo{author}{\bibfnamefont{D.}~\bibnamefont{Wortmann}},
  \bibinfo{author}{\bibfnamefont{S.}~\bibnamefont{Heinze}},
  \bibinfo{author}{\bibfnamefont{P.}~\bibnamefont{Kurz}},
  \bibinfo{author}{\bibfnamefont{G.}~\bibnamefont{Bihlmayer}},
  \bibnamefont{and} \bibinfo{author}{\bibfnamefont{S.}~\bibnamefont{Bl\"ugel}},
  ``Resolving complex atomic-scale spin structures by spin-polarized scanning
  tunneling microscopy,'' \bibinfo{journal}{Phys. Rev. Lett.}
  \textbf{\bibinfo{volume}{86}}, \bibinfo{pages}{4132} (\bibinfo{year}{2001}).

\bibitem[{\citenamefont{Burtzlaff et~al.}(2015)\citenamefont{Burtzlaff,
  Weismann, Brandbyge, and Berndt}}]{Burtzlaff2015}
\bibinfo{author}{\bibfnamefont{A.}~\bibnamefont{Burtzlaff}},
  \bibinfo{author}{\bibfnamefont{A.}~\bibnamefont{Weismann}},
  \bibinfo{author}{\bibfnamefont{M.}~\bibnamefont{Brandbyge}},
  \bibnamefont{and} \bibinfo{author}{\bibfnamefont{R.}~\bibnamefont{Berndt}},
  ``Shot noise as a probe of spin-polarized transport through single atoms,''
  \bibinfo{journal}{Phys. Rev. Lett.} \textbf{\bibinfo{volume}{114}},
  \bibinfo{pages}{016602} (\bibinfo{year}{2015}).

\bibitem[{\citenamefont{B\"uttiker et~al.}(1985)\citenamefont{B\"uttiker, Imry,
  Landauer, and Pinhas}}]{Buettiker1985}
\bibinfo{author}{\bibfnamefont{M.}~\bibnamefont{B\"uttiker}},
  \bibinfo{author}{\bibfnamefont{Y.}~\bibnamefont{Imry}},
  \bibinfo{author}{\bibfnamefont{R.}~\bibnamefont{Landauer}}, \bibnamefont{and}
  \bibinfo{author}{\bibfnamefont{S.}~\bibnamefont{Pinhas}}, ``Generalized
  many-channel conductance formula with application to small rings,''
  \bibinfo{journal}{Phys. Rev. B} \textbf{\bibinfo{volume}{31}},
  \bibinfo{pages}{6207} (\bibinfo{year}{1985}).

\bibitem[{\citenamefont{Areshkin and White}(2007)}]{Areshkin2007}
\bibinfo{author}{\bibfnamefont{D.~A.} \bibnamefont{Areshkin}} \bibnamefont{and}
  \bibinfo{author}{\bibfnamefont{C.~T.} \bibnamefont{White}}, ``Building blocks
  for integrated graphene circuits,'' \bibinfo{journal}{Nano Lett.}
  \textbf{\bibinfo{volume}{7}}, \bibinfo{pages}{3253} (\bibinfo{year}{2007}).

\bibitem[{\citenamefont{Jayasekera and Mintmire}(2007)}]{Jayasekera2007}
\bibinfo{author}{\bibfnamefont{T.}~\bibnamefont{Jayasekera}} \bibnamefont{and}
  \bibinfo{author}{\bibfnamefont{J.~W.} \bibnamefont{Mintmire}}, ``Transport in
  multiterminal graphene nanodevices,'' \bibinfo{journal}{Nanotechn.}
  \textbf{\bibinfo{volume}{18}}, \bibinfo{pages}{424033}
  (\bibinfo{year}{2007}).

\bibitem[{\citenamefont{Botello-Méndez
  et~al.}(2011)\citenamefont{Botello-Méndez, Cruz-Silva, Romo-Herrera,
  López-Urías, Terrones, Sumpter, Terrones, Charlier, and
  Meunier}}]{Botello-Mendez2011}
\bibinfo{author}{\bibfnamefont{A.~R.} \bibnamefont{Botello-Méndez}},
  \bibinfo{author}{\bibfnamefont{E.}~\bibnamefont{Cruz-Silva}},
  \bibinfo{author}{\bibfnamefont{J.~M.} \bibnamefont{Romo-Herrera}},
  \bibinfo{author}{\bibfnamefont{F.}~\bibnamefont{López-Urías}},
  \bibinfo{author}{\bibfnamefont{M.}~\bibnamefont{Terrones}},
  \bibinfo{author}{\bibfnamefont{B.~G.} \bibnamefont{Sumpter}},
  \bibinfo{author}{\bibfnamefont{H.}~\bibnamefont{Terrones}},
  \bibinfo{author}{\bibfnamefont{J.-C.} \bibnamefont{Charlier}},
  \bibnamefont{and} \bibinfo{author}{\bibfnamefont{V.}~\bibnamefont{Meunier}},
  ``Quantum transport in graphene nanonetworks,'' \bibinfo{journal}{Nano Lett.}
  \textbf{\bibinfo{volume}{11}}, \bibinfo{pages}{3058} (\bibinfo{year}{2011}).

\bibitem[{\citenamefont{Cary et~al.}(2014)\citenamefont{Cary, Costa Gir\~ao,
  and Meunier}}]{CaCoMe.14.Electronicpropertiesthree}
\bibinfo{author}{\bibfnamefont{T.}~\bibnamefont{Cary}},
  \bibinfo{author}{\bibfnamefont{E.}~\bibnamefont{Costa Gir\~ao}},
  \bibnamefont{and} \bibinfo{author}{\bibfnamefont{V.}~\bibnamefont{Meunier}},
  ``Electronic properties of three-terminal graphitic nanowiggles,''
  \bibinfo{journal}{Phys. Rev. B} \textbf{\bibinfo{volume}{90}},
  \bibinfo{pages}{115409} (\bibinfo{year}{2014}).

\bibitem[{\citenamefont{Lima et~al.}(2016)\citenamefont{Lima, Hern{\'{a}}ndez,
  Pinheiro, and Lewenkopf}}]{Lima2016}
\bibinfo{author}{\bibfnamefont{L.~R.~F.} \bibnamefont{Lima}},
  \bibinfo{author}{\bibfnamefont{A.~R.} \bibnamefont{Hern{\'{a}}ndez}},
  \bibinfo{author}{\bibfnamefont{F.~A.} \bibnamefont{Pinheiro}},
  \bibnamefont{and}
  \bibinfo{author}{\bibfnamefont{C.}~\bibnamefont{Lewenkopf}}, ``A 50/50
  electronic beam splitter in graphene nanoribbons as a building block for
  electron optics,'' \bibinfo{journal}{J.~Phys.: Condens.~Matter}
  \textbf{\bibinfo{volume}{28}}, \bibinfo{pages}{505303}
  (\bibinfo{year}{2016}).

\bibitem[{\citenamefont{Brandimarte et~al.}(2017)\citenamefont{Brandimarte,
  Engelund, Papior, Garcia-Lekue, Frederiksen, and
  S\'anchez-Portal}}]{Brandimarte2017}
\bibinfo{author}{\bibfnamefont{P.}~\bibnamefont{Brandimarte}},
  \bibinfo{author}{\bibfnamefont{M.}~\bibnamefont{Engelund}},
  \bibinfo{author}{\bibfnamefont{N.}~\bibnamefont{Papior}},
  \bibinfo{author}{\bibfnamefont{A.}~\bibnamefont{Garcia-Lekue}},
  \bibinfo{author}{\bibfnamefont{T.}~\bibnamefont{Frederiksen}},
  \bibnamefont{and}
  \bibinfo{author}{\bibfnamefont{D.}~\bibnamefont{S\'anchez-Portal}}, ``A
  tunable electronic beam splitter realized with crossed graphene
  nanoribbons,'' \bibinfo{journal}{J.~Chem.~Phys.}
  \textbf{\bibinfo{volume}{146}}, \bibinfo{pages}{092318}
  (\bibinfo{year}{2017}).

\bibitem[{\citenamefont{Sanz et~al.}(2020)\citenamefont{Sanz, Brandimarte,
  Giedke, S\'anchez-Portal, and Frederiksen}}]{Sanz2020}
\bibinfo{author}{\bibfnamefont{S.}~\bibnamefont{Sanz}},
  \bibinfo{author}{\bibfnamefont{P.}~\bibnamefont{Brandimarte}},
  \bibinfo{author}{\bibfnamefont{G.}~\bibnamefont{Giedke}},
  \bibinfo{author}{\bibfnamefont{D.}~\bibnamefont{S\'anchez-Portal}},
  \bibnamefont{and}
  \bibinfo{author}{\bibfnamefont{T.}~\bibnamefont{Frederiksen}}, ``Crossed
  graphene nanoribbons as beam splitters and mirrors for electron quantum
  optics,'' \bibinfo{journal}{Phys. Rev. B} \textbf{\bibinfo{volume}{102}},
  \bibinfo{pages}{035436} (\bibinfo{year}{2020}).

\bibitem[{\citenamefont{Brey and Fertig}(2006)}]{Brey2006}
\bibinfo{author}{\bibfnamefont{L.}~\bibnamefont{Brey}} \bibnamefont{and}
  \bibinfo{author}{\bibfnamefont{H.~A.} \bibnamefont{Fertig}}, ``Electronic
  states of graphene nanoribbons studied with the {Dirac} equation,''
  \bibinfo{journal}{Phys. Rev. B} \textbf{\bibinfo{volume}{73}},
  \bibinfo{pages}{235411} (\bibinfo{year}{2006}).

\bibitem[{\citenamefont{Wakabayashi et~al.}(2007)\citenamefont{Wakabayashi,
  Takane, and Sigrist}}]{Wakabayashi2007}
\bibinfo{author}{\bibfnamefont{K.}~\bibnamefont{Wakabayashi}},
  \bibinfo{author}{\bibfnamefont{Y.}~\bibnamefont{Takane}}, \bibnamefont{and}
  \bibinfo{author}{\bibfnamefont{M.}~\bibnamefont{Sigrist}}, ``Perfectly
  conducting channel and universality crossover in disordered graphene
  nanoribbons,'' \bibinfo{journal}{Phys. Rev. Lett.}
  \textbf{\bibinfo{volume}{99}}, \bibinfo{pages}{036601}
  (\bibinfo{year}{2007}).

\bibitem[{\citenamefont{Fujita et~al.}(1996)\citenamefont{Fujita, Wakabayashi,
  Nakada, and Kusakabe}}]{Fujita1996}
\bibinfo{author}{\bibfnamefont{M.}~\bibnamefont{Fujita}},
  \bibinfo{author}{\bibfnamefont{K.}~\bibnamefont{Wakabayashi}},
  \bibinfo{author}{\bibfnamefont{K.}~\bibnamefont{Nakada}}, \bibnamefont{and}
  \bibinfo{author}{\bibfnamefont{K.}~\bibnamefont{Kusakabe}}, ``Peculiar
  localized state at zigzag graphite edge,'' \bibinfo{journal}{J. Phys. Soc.
  Jpn.} \textbf{\bibinfo{volume}{65}}, \bibinfo{pages}{1920}
  (\bibinfo{year}{1996}).

\bibitem[{\citenamefont{Blackwell et~al.}(2021)\citenamefont{Blackwell, Zhao,
  Brooks, Zhu, Piskun, Wang, Delgado, Lee, Louie, and
  Fischer}}]{BlZhBr.21.Spinsplittingdopant}
\bibinfo{author}{\bibfnamefont{R.~E.} \bibnamefont{Blackwell}},
  \bibinfo{author}{\bibfnamefont{F.}~\bibnamefont{Zhao}},
  \bibinfo{author}{\bibfnamefont{E.}~\bibnamefont{Brooks}},
  \bibinfo{author}{\bibfnamefont{J.}~\bibnamefont{Zhu}},
  \bibinfo{author}{\bibfnamefont{I.}~\bibnamefont{Piskun}},
  \bibinfo{author}{\bibfnamefont{S.}~\bibnamefont{Wang}},
  \bibinfo{author}{\bibfnamefont{A.}~\bibnamefont{Delgado}},
  \bibinfo{author}{\bibfnamefont{Y.-L.} \bibnamefont{Lee}},
  \bibinfo{author}{\bibfnamefont{S.~G.} \bibnamefont{Louie}}, \bibnamefont{and}
  \bibinfo{author}{\bibfnamefont{F.~R.} \bibnamefont{Fischer}}, ``Spin
  splitting of dopant edge state in magnetic zigzag graphene nanoribbons,''
  \bibinfo{journal}{Nature} \textbf{\bibinfo{volume}{600}},
  \bibinfo{pages}{647} (\bibinfo{year}{2021}).

\bibitem[{\citenamefont{Wimmer et~al.}(2008)\citenamefont{Wimmer, Adagideli,
  Berber, Tom\'anek, and Richter}}]{WiAdBe.08.SpinCurrentsRough}
\bibinfo{author}{\bibfnamefont{M.}~\bibnamefont{Wimmer}},
  \bibinfo{author}{\bibfnamefont{I.}~\bibnamefont{Adagideli}},
  \bibinfo{author}{\bibfnamefont{S.}~\bibnamefont{Berber}},
  \bibinfo{author}{\bibfnamefont{D.}~\bibnamefont{Tom\'anek}},
  \bibnamefont{and} \bibinfo{author}{\bibfnamefont{K.}~\bibnamefont{Richter}},
  ``Spin currents in rough graphene nanoribbons: Universal fluctuations and
  spin injection,'' \bibinfo{journal}{Phys. Rev. Lett.}
  \textbf{\bibinfo{volume}{100}}, \bibinfo{pages}{177207}
  (\bibinfo{year}{2008}).

\bibitem[{\citenamefont{Sanz et~al.}(2021)\citenamefont{Sanz, Papior,
  Brandbyge, and Frederiksen}}]{dipc_hubbard}
\bibinfo{author}{\bibfnamefont{S.}~\bibnamefont{Sanz}},
  \bibinfo{author}{\bibfnamefont{N.}~\bibnamefont{Papior}},
  \bibinfo{author}{\bibfnamefont{M.}~\bibnamefont{Brandbyge}},
  \bibnamefont{and}
  \bibinfo{author}{\bibfnamefont{T.}~\bibnamefont{Frederiksen}}, ``hubbard:
  v0.1.0,'' (\bibinfo{year}{2021}).

\bibitem[{\citenamefont{Hubbard}(1963)}]{Hubbard1963}
\bibinfo{author}{\bibfnamefont{J.}~\bibnamefont{Hubbard}}, ``Electron
  correlations in narrow energy bands,'' \bibinfo{journal}{Proc. R. Soc. A}
  \textbf{\bibinfo{volume}{276}}, \bibinfo{pages}{238} (\bibinfo{year}{1963}).

\bibitem[{\citenamefont{Fern\'andez-Rossier and
  Palacios}(2007)}]{FePa.07.MagnetismGrapheneNanoislands}
\bibinfo{author}{\bibfnamefont{J.}~\bibnamefont{Fern\'andez-Rossier}}
  \bibnamefont{and} \bibinfo{author}{\bibfnamefont{J.~J.}
  \bibnamefont{Palacios}}, ``Magnetism in graphene nanoislands,''
  \bibinfo{journal}{Phys. Rev. Lett.} \textbf{\bibinfo{volume}{99}},
  \bibinfo{pages}{177204} (\bibinfo{year}{2007}).

\bibitem[{\citenamefont{Li et~al.}(2019)\citenamefont{Li, Sanz, Corso, Choi,
  Pe{\~{n}}a, Frederiksen, and Pascual}}]{Li2019}
\bibinfo{author}{\bibfnamefont{J.}~\bibnamefont{Li}},
  \bibinfo{author}{\bibfnamefont{S.}~\bibnamefont{Sanz}},
  \bibinfo{author}{\bibfnamefont{M.}~\bibnamefont{Corso}},
  \bibinfo{author}{\bibfnamefont{D.~J.} \bibnamefont{Choi}},
  \bibinfo{author}{\bibfnamefont{D.}~\bibnamefont{Pe{\~{n}}a}},
  \bibinfo{author}{\bibfnamefont{T.}~\bibnamefont{Frederiksen}},
  \bibnamefont{and} \bibinfo{author}{\bibfnamefont{J.~I.}
  \bibnamefont{Pascual}}, ``Single spin localization and manipulation in
  graphene open-shell nanostructures,'' \bibinfo{journal}{Nat.~Commun.}
  \textbf{\bibinfo{volume}{10}}, \bibinfo{pages}{200} (\bibinfo{year}{2019}).

\bibitem[{\citenamefont{Papior et~al.}(2017)\citenamefont{Papior, Lorente,
  Frederiksen, Garc\'ia, and Brandbyge}}]{Papior2017}
\bibinfo{author}{\bibfnamefont{N.}~\bibnamefont{Papior}},
  \bibinfo{author}{\bibfnamefont{N.}~\bibnamefont{Lorente}},
  \bibinfo{author}{\bibfnamefont{T.}~\bibnamefont{Frederiksen}},
  \bibinfo{author}{\bibfnamefont{A.}~\bibnamefont{Garc\'ia}}, \bibnamefont{and}
  \bibinfo{author}{\bibfnamefont{M.}~\bibnamefont{Brandbyge}}, ``Improvements
  on non-equilibrium and transport green function techniques: The
  next-generation {TranSiesta},'' \bibinfo{journal}{Comp. Phys. Commun.}
  \textbf{\bibinfo{volume}{212}}, \bibinfo{pages}{8 } (\bibinfo{year}{2017}).

\bibitem[{\citenamefont{Papior}(2021)}]{zerothi_sisl}
\bibinfo{author}{\bibfnamefont{N.}~\bibnamefont{Papior}}, ``sisl: v0.11.0,''
  (\bibinfo{year}{2021}).

\bibitem[{\citenamefont{{See {Supplemental Material} [url] for details on
  methodology and additional calculations, which includes
  {Refs.~[52-58]}}}()}]{SM}
\bibinfo{author}{\bibnamefont{{See {Supplemental Material} [url] for details on
  methodology and additional calculations, which includes {Refs.~[52-58]}}}}.

\bibitem[{\citenamefont{Z{\^{a}}rbo and Nikoli{\'{c}}}(2007)}]{Zarbo2007}
\bibinfo{author}{\bibfnamefont{L.~P.} \bibnamefont{Z{\^{a}}rbo}}
  \bibnamefont{and} \bibinfo{author}{\bibfnamefont{B.~K.}
  \bibnamefont{Nikoli{\'{c}}}}, ``Spatial distribution of local currents of
  massless {Dirac} fermions in quantum transport through graphene
  nanoribbons,'' \bibinfo{journal}{Europhys. Lett. ({EPL})}
  \textbf{\bibinfo{volume}{80}}, \bibinfo{pages}{47001} (\bibinfo{year}{2007}).

\bibitem[{\citenamefont{Loth et~al.}(2010)\citenamefont{Loth, von Bergmann,
  Ternes, Otte, Lutz, and Heinrich}}]{LoBeTe.10.Controllingstateof}
\bibinfo{author}{\bibfnamefont{S.}~\bibnamefont{Loth}},
  \bibinfo{author}{\bibfnamefont{K.}~\bibnamefont{von Bergmann}},
  \bibinfo{author}{\bibfnamefont{M.}~\bibnamefont{Ternes}},
  \bibinfo{author}{\bibfnamefont{A.~F.} \bibnamefont{Otte}},
  \bibinfo{author}{\bibfnamefont{C.~P.} \bibnamefont{Lutz}}, \bibnamefont{and}
  \bibinfo{author}{\bibfnamefont{A.~J.} \bibnamefont{Heinrich}}, ``Controlling
  the state of quantum spins with electric currents,''
  \bibinfo{journal}{Nat.~Phys.} \textbf{\bibinfo{volume}{6}},
  \bibinfo{pages}{340} (\bibinfo{year}{2010}).

\bibitem[{\citenamefont{Bocquillon et~al.}(2014)\citenamefont{Bocquillon,
  Freulon, Parmentier, Berroir, Plaçais, Wahl, Rech, Jonckheere, Martin,
  Grenier et~al.}}]{BoFrPa.14.Electronquantumoptics}
\bibinfo{author}{\bibfnamefont{E.}~\bibnamefont{Bocquillon}},
  \bibinfo{author}{\bibfnamefont{V.}~\bibnamefont{Freulon}},
  \bibinfo{author}{\bibfnamefont{F.~D.} \bibnamefont{Parmentier}},
  \bibinfo{author}{\bibfnamefont{J.-M.} \bibnamefont{Berroir}},
  \bibinfo{author}{\bibfnamefont{B.}~\bibnamefont{Plaçais}},
  \bibinfo{author}{\bibfnamefont{C.}~\bibnamefont{Wahl}},
  \bibinfo{author}{\bibfnamefont{J.}~\bibnamefont{Rech}},
  \bibinfo{author}{\bibfnamefont{T.}~\bibnamefont{Jonckheere}},
  \bibinfo{author}{\bibfnamefont{T.}~\bibnamefont{Martin}},
  \bibinfo{author}{\bibfnamefont{C.}~\bibnamefont{Grenier}},
  \bibnamefont{et~al.}, ``Electron quantum optics in ballistic chiral
  conductors,'' \bibinfo{journal}{Ann.~Phys.} \textbf{\bibinfo{volume}{526}},
  \bibinfo{pages}{1} (\bibinfo{year}{2014}).

\bibitem[{\citenamefont{Jacobse et~al.}(2018)\citenamefont{Jacobse, Mangnus,
  Zevenhuizen, and Swart}}]{JaMaZe.18.MappingConductanceElectronically}
\bibinfo{author}{\bibfnamefont{P.}~\bibnamefont{Jacobse}},
  \bibinfo{author}{\bibfnamefont{M.~J.~J.} \bibnamefont{Mangnus}},
  \bibinfo{author}{\bibfnamefont{S.~J.~M.} \bibnamefont{Zevenhuizen}},
  \bibnamefont{and} \bibinfo{author}{\bibfnamefont{I.}~\bibnamefont{Swart}},
  ``Mapping the conductance of electronically decoupled graphene nanoribbons,''
  \bibinfo{journal}{ACS Nano} \textbf{\bibinfo{volume}{12}},
  \bibinfo{pages}{7048} (\bibinfo{year}{2018}).

\bibitem[{\citenamefont{Baumann et~al.}(2015)\citenamefont{Baumann, Paul, Choi,
  Lutz, Ardavan, and Heinrich}}]{BaPaCh.15.Electronparamagneticresonance}
\bibinfo{author}{\bibfnamefont{S.}~\bibnamefont{Baumann}},
  \bibinfo{author}{\bibfnamefont{W.}~\bibnamefont{Paul}},
  \bibinfo{author}{\bibfnamefont{T.}~\bibnamefont{Choi}},
  \bibinfo{author}{\bibfnamefont{C.~P.} \bibnamefont{Lutz}},
  \bibinfo{author}{\bibfnamefont{A.}~\bibnamefont{Ardavan}}, \bibnamefont{and}
  \bibinfo{author}{\bibfnamefont{A.~J.} \bibnamefont{Heinrich}}, ``Electron
  paramagnetic resonance of individual atoms on a surface,''
  \bibinfo{journal}{Science} \textbf{\bibinfo{volume}{350}},
  \bibinfo{pages}{417} (\bibinfo{year}{2015}).

\bibitem[{\citenamefont{Ammerman et~al.}(2021)\citenamefont{Ammerman, Jelic,
  Wei, Breslin, Hassan, Everett, Lee, Sun, Pignedoli, Ruffieux
  et~al.}}]{AmJeWe.21.Lightwavedrivenscanning}
\bibinfo{author}{\bibfnamefont{S.~E.} \bibnamefont{Ammerman}},
  \bibinfo{author}{\bibfnamefont{V.}~\bibnamefont{Jelic}},
  \bibinfo{author}{\bibfnamefont{Y.}~\bibnamefont{Wei}},
  \bibinfo{author}{\bibfnamefont{V.~N.} \bibnamefont{Breslin}},
  \bibinfo{author}{\bibfnamefont{M.}~\bibnamefont{Hassan}},
  \bibinfo{author}{\bibfnamefont{N.}~\bibnamefont{Everett}},
  \bibinfo{author}{\bibfnamefont{S.}~\bibnamefont{Lee}},
  \bibinfo{author}{\bibfnamefont{Q.}~\bibnamefont{Sun}},
  \bibinfo{author}{\bibfnamefont{C.~A.} \bibnamefont{Pignedoli}},
  \bibinfo{author}{\bibfnamefont{P.}~\bibnamefont{Ruffieux}},
  \bibnamefont{et~al.}, ``Lightwave-driven scanning tunnelling spectroscopy of
  atomically precise graphene nanoribbons,'' \bibinfo{journal}{Nat. Commun.}
  \textbf{\bibinfo{volume}{12}}, \bibinfo{pages}{6794} (\bibinfo{year}{2021}).

\bibitem[{\citenamefont{Kolmer et~al.}(2019)\citenamefont{Kolmer, Brandimarte,
  Lis, Zuzak, Godlewski, Kawai, Garcia-Lekue, Lorente, Frederiksen, Joachim
  et~al.}}]{KoBrLi.19.Electronictransportplanar}
\bibinfo{author}{\bibfnamefont{M.}~\bibnamefont{Kolmer}},
  \bibinfo{author}{\bibfnamefont{P.}~\bibnamefont{Brandimarte}},
  \bibinfo{author}{\bibfnamefont{J.}~\bibnamefont{Lis}},
  \bibinfo{author}{\bibfnamefont{R.}~\bibnamefont{Zuzak}},
  \bibinfo{author}{\bibfnamefont{S.}~\bibnamefont{Godlewski}},
  \bibinfo{author}{\bibfnamefont{H.}~\bibnamefont{Kawai}},
  \bibinfo{author}{\bibfnamefont{A.}~\bibnamefont{Garcia-Lekue}},
  \bibinfo{author}{\bibfnamefont{N.}~\bibnamefont{Lorente}},
  \bibinfo{author}{\bibfnamefont{T.}~\bibnamefont{Frederiksen}},
  \bibinfo{author}{\bibfnamefont{C.}~\bibnamefont{Joachim}},
  \bibnamefont{et~al.}, ``Electronic transport in planar atomic-scale
  structures measured by two-probe scanning tunneling spectroscopy,''
  \bibinfo{journal}{Nat. Commun.} \textbf{\bibinfo{volume}{10}},
  \bibinfo{pages}{1573} (\bibinfo{year}{2019}).

\bibitem[{\citenamefont{Chong et~al.}(2018)\citenamefont{Chong, Afshar-Imani,
  Scheurer, Cardoso, Ferretti, Prezzi, and
  Schull}}]{ChAfSc.18.BrightElectroluminescenceSingle}
\bibinfo{author}{\bibfnamefont{M.~C.} \bibnamefont{Chong}},
  \bibinfo{author}{\bibfnamefont{N.}~\bibnamefont{Afshar-Imani}},
  \bibinfo{author}{\bibfnamefont{F.}~\bibnamefont{Scheurer}},
  \bibinfo{author}{\bibfnamefont{C.}~\bibnamefont{Cardoso}},
  \bibinfo{author}{\bibfnamefont{A.}~\bibnamefont{Ferretti}},
  \bibinfo{author}{\bibfnamefont{D.}~\bibnamefont{Prezzi}}, \bibnamefont{and}
  \bibinfo{author}{\bibfnamefont{G.}~\bibnamefont{Schull}}, ``Bright
  electroluminescence from single graphene nanoribbon junctions,''
  \bibinfo{journal}{Nano Lett.} \textbf{\bibinfo{volume}{18}},
  \bibinfo{pages}{175} (\bibinfo{year}{2018}).

\bibitem[{\citenamefont{Beenakker et~al.}(2004)\citenamefont{Beenakker,
  DiVincenzo, Emary, and Kindermann}}]{BeDiEm.04.ChargeDetectionEnables}
\bibinfo{author}{\bibfnamefont{C.~W.~J.} \bibnamefont{Beenakker}},
  \bibinfo{author}{\bibfnamefont{D.~P.} \bibnamefont{DiVincenzo}},
  \bibinfo{author}{\bibfnamefont{C.}~\bibnamefont{Emary}}, \bibnamefont{and}
  \bibinfo{author}{\bibfnamefont{M.}~\bibnamefont{Kindermann}}, ``Charge
  detection enables free-electron quantum computation,''
  \bibinfo{journal}{Phys. Rev. Lett.} \textbf{\bibinfo{volume}{93}},
  \bibinfo{pages}{020501} (\bibinfo{year}{2004}).

\bibitem[{\citenamefont{Mazza et~al.}(2013)\citenamefont{Mazza, Braunecker,
  Recher, and Levy~Yeyati}}]{MaBrRe.13.Spinfilteringentanglement}
\bibinfo{author}{\bibfnamefont{F.}~\bibnamefont{Mazza}},
  \bibinfo{author}{\bibfnamefont{B.}~\bibnamefont{Braunecker}},
  \bibinfo{author}{\bibfnamefont{P.}~\bibnamefont{Recher}}, \bibnamefont{and}
  \bibinfo{author}{\bibfnamefont{A.}~\bibnamefont{Levy~Yeyati}}, ``Spin
  filtering and entanglement detection due to spin-orbit interaction in carbon
  nanotube cross-junctions,'' \bibinfo{journal}{Phys. Rev. B}
  \textbf{\bibinfo{volume}{88}}, \bibinfo{pages}{195403}
  (\bibinfo{year}{2013}).

\bibitem[{\citenamefont{Feldner et~al.}(2011)\citenamefont{Feldner, Meng, Lang,
  Assaad, Wessel, and Honecker}}]{FeMeLa.11.DynamicalSignaturesEdge}
\bibinfo{author}{\bibfnamefont{H.}~\bibnamefont{Feldner}},
  \bibinfo{author}{\bibfnamefont{Z.~Y.} \bibnamefont{Meng}},
  \bibinfo{author}{\bibfnamefont{T.~C.} \bibnamefont{Lang}},
  \bibinfo{author}{\bibfnamefont{F.~F.} \bibnamefont{Assaad}},
  \bibinfo{author}{\bibfnamefont{S.}~\bibnamefont{Wessel}}, \bibnamefont{and}
  \bibinfo{author}{\bibfnamefont{A.}~\bibnamefont{Honecker}}, ``Dynamical
  signatures of edge-state magnetism on graphene nanoribbons,''
  \bibinfo{journal}{Phys. Rev. Lett.} \textbf{\bibinfo{volume}{106}},
  \bibinfo{pages}{226401} (\bibinfo{year}{2011}).

\bibitem[{\citenamefont{Brandbyge et~al.}(2002)\citenamefont{Brandbyge, Mozos,
  Ordej\'on, Taylor, and Stokbro}}]{Brandbyge2002}
\bibinfo{author}{\bibfnamefont{M.}~\bibnamefont{Brandbyge}},
  \bibinfo{author}{\bibfnamefont{J.-L.} \bibnamefont{Mozos}},
  \bibinfo{author}{\bibfnamefont{P.}~\bibnamefont{Ordej\'on}},
  \bibinfo{author}{\bibfnamefont{J.}~\bibnamefont{Taylor}}, \bibnamefont{and}
  \bibinfo{author}{\bibfnamefont{K.}~\bibnamefont{Stokbro}},
  ``Density-functional method for nonequilibrium electron transport,''
  \bibinfo{journal}{Phys. Rev. B} \textbf{\bibinfo{volume}{65}},
  \bibinfo{pages}{165401} (\bibinfo{year}{2002}).

\bibitem[{\citenamefont{Sancho et~al.}(1985)\citenamefont{Sancho, Sancho,
  Sancho, and Rubio}}]{Sancho1985}
\bibinfo{author}{\bibfnamefont{M.~P.~L.} \bibnamefont{Sancho}},
  \bibinfo{author}{\bibfnamefont{J.~M.~L.} \bibnamefont{Sancho}},
  \bibinfo{author}{\bibfnamefont{J.~M.~L.} \bibnamefont{Sancho}},
  \bibnamefont{and} \bibinfo{author}{\bibfnamefont{J.}~\bibnamefont{Rubio}},
  ``Highly convergent schemes for the calculation of bulk and surface {Green}
  functions,'' \bibinfo{journal}{J.~Phys.~F:~Met.~Phys.}
  \textbf{\bibinfo{volume}{15}}, \bibinfo{pages}{851} (\bibinfo{year}{1985}).

\bibitem[{\citenamefont{Asano and Nakamura}(2019)}]{Asano2019}
\bibinfo{author}{\bibfnamefont{T.}~\bibnamefont{Asano}} \bibnamefont{and}
  \bibinfo{author}{\bibfnamefont{J.}~\bibnamefont{Nakamura}},
  ``Edge-state-induced stacking of zigzag graphene nanoribbons,''
  \bibinfo{journal}{ACS Omega} \textbf{\bibinfo{volume}{4}},
  \bibinfo{pages}{22035} (\bibinfo{year}{2019}).

\bibitem[{\citenamefont{Mostaani et~al.}(2015)\citenamefont{Mostaani, Drummond,
  and Fal'ko}}]{Mostaani2015}
\bibinfo{author}{\bibfnamefont{E.}~\bibnamefont{Mostaani}},
  \bibinfo{author}{\bibfnamefont{N.~D.} \bibnamefont{Drummond}},
  \bibnamefont{and} \bibinfo{author}{\bibfnamefont{V.~I.}
  \bibnamefont{Fal'ko}}, ``Quantum {Monte Carlo} calculation of the binding
  energy of bilayer graphene,'' \bibinfo{journal}{Phys. Rev. Lett.}
  \textbf{\bibinfo{volume}{115}}, \bibinfo{pages}{115501}
  (\bibinfo{year}{2015}).

\bibitem[{\citenamefont{Lee et~al.}(2005)\citenamefont{Lee, Son, Park, Han, and
  Yu}}]{Lee2005}
\bibinfo{author}{\bibfnamefont{H.}~\bibnamefont{Lee}},
  \bibinfo{author}{\bibfnamefont{Y.-W.} \bibnamefont{Son}},
  \bibinfo{author}{\bibfnamefont{N.}~\bibnamefont{Park}},
  \bibinfo{author}{\bibfnamefont{S.}~\bibnamefont{Han}}, \bibnamefont{and}
  \bibinfo{author}{\bibfnamefont{J.}~\bibnamefont{Yu}}, ``Magnetic ordering at
  the edges of graphitic fragments: Magnetic tail interactions between the
  edge-localized states,'' \bibinfo{journal}{Phys. Rev. B}
  \textbf{\bibinfo{volume}{72}}, \bibinfo{pages}{174431}
  (\bibinfo{year}{2005}).

\bibitem[{\citenamefont{Cahay et~al.}(1988)\citenamefont{Cahay, McLennan, and
  Datta}}]{Cahay1988}
\bibinfo{author}{\bibfnamefont{M.}~\bibnamefont{Cahay}},
  \bibinfo{author}{\bibfnamefont{M.}~\bibnamefont{McLennan}}, \bibnamefont{and}
  \bibinfo{author}{\bibfnamefont{S.}~\bibnamefont{Datta}}, ``Conductance of an
  array of elastic scatterers: A scattering-matrix approach,''
  \bibinfo{journal}{Phys. Rev. B} \textbf{\bibinfo{volume}{37}},
  \bibinfo{pages}{10125} (\bibinfo{year}{1988}).

\end{thebibliography}

\includepdf[pages={{},-}, pagecommand={\clearpage \thispagestyle{empty}}, scale=1]{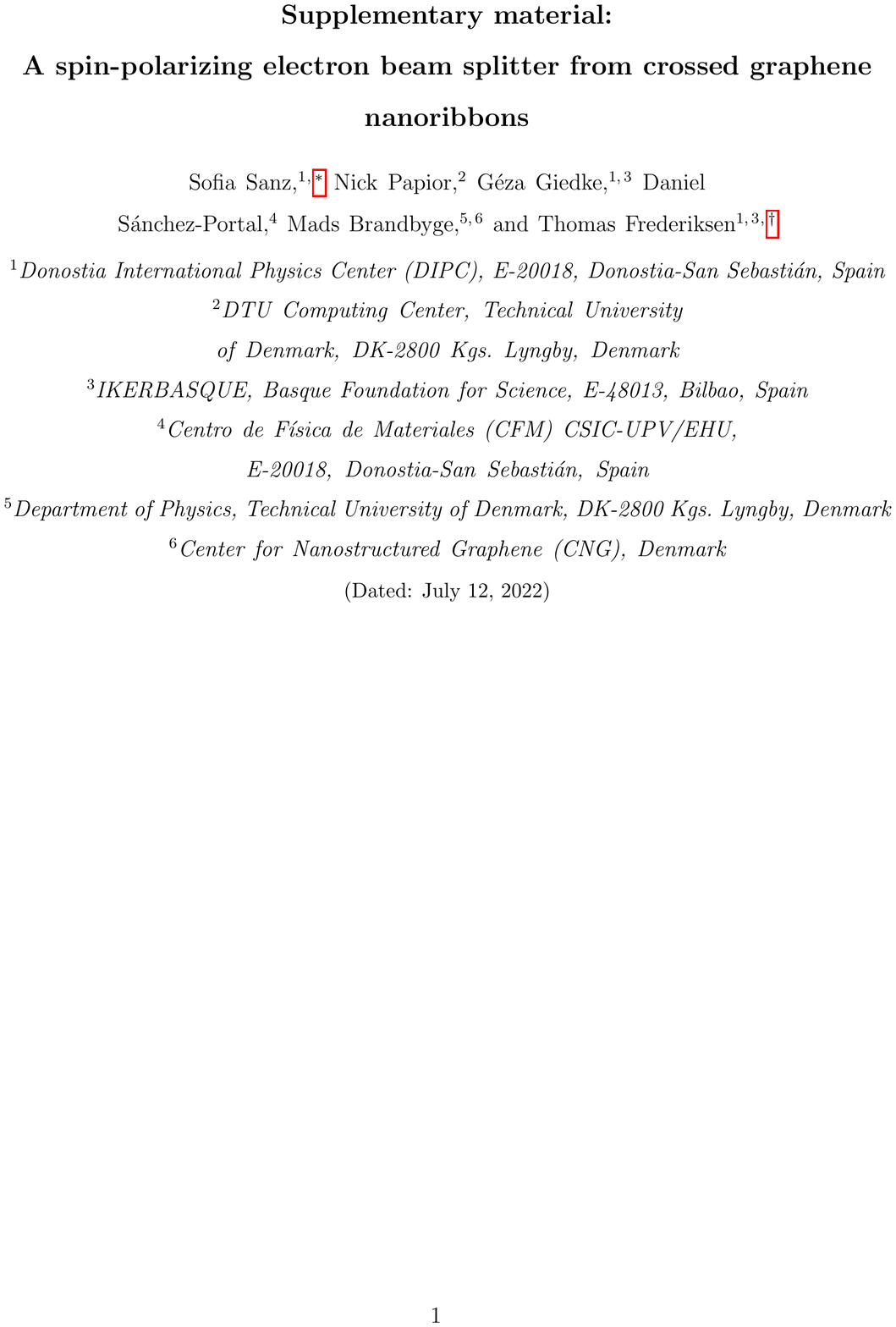}
\thispagestyle{empty}

\end{document}